\documentstyle[12pt]{article}
\newcommand{\beq}{\begin{equation}}
\newcommand{\eeq}{\end{equation}}
\newcommand{\bea}{\begin{eqnarray}}
\newcommand{\eea}{\end{eqnarray}}

\newcommand{\th}{\theta}

\newcommand{\be}{\beta}

\newcommand{\kb}{\bar{k}}

\begin{document}
\setcounter{page}{0}
\topmargin 0pt
\oddsidemargin 5mm
\renewcommand{\thefootnote}{\fnsymbol{footnote}}
\newpage
\setcounter{page}{0}
\begin{titlepage}
\begin{flushright}
USP-IFQSC/TH/93-07
\end{flushright}
\vspace{0.5cm}
\begin{center}
{\large {\bf Form Factors in  Affine Toda Field Theories}} \\
\vspace{1.8cm}
{\large Andreas Fring
\footnote{To appear in the Proceedings of the VII Andr\'e Swieca Summer
School, Campos do Jord\~ao, Brasil, 1993 }}\\
\vspace{0.5cm}
{\em Universidade de S\~ao Paulo, \\
Caixa Postal 369, CEP 13560 S\~ao Carlos-SP, Brasil}\\
\vspace{3cm}
\renewcommand{\thefootnote}{\arabic{footnote}}
\setcounter{footnote}{0}
\begin{abstract}
{I briefly review the properties of classical affine Toda field
theories and indicate how some of this features survive in the quantum
theory on-shell. I demonstrate how this knowledge can be extended off-shell,
i.e. how to compute correlation functions for completely integrable models
via the form factor approach. For the latter I present an axiomatic system
and two explicit computation (the Sinh-Gordon theory and the Bullough-Dodd
model) which provide a consistent solution of it.}
\end{abstract}
\vspace{.3cm}
\centerline{March 1993}
 \end{center}
\end{titlepage}
\newpage
\section{Introduction}

Completely integrable models have attracted a lot of attention and
research activity in recent years, since they offer the possibility to
obtain exact, i.e. non-perturbative, results in quantum field theory.
Affine Toda Field Theories (ATFTs) represent a very important example of
this models, since they are explicit Lagrangian  versions of integrable
deformations of conformal field theories. The breaking of the conformal
symmetry simply corresponds to an affinisation of the Lie algebra {\bf g}
underlying the conformally invariant theory. Many other features of ATFT
can be expressed as well very neatly in terms of Lie algebraic quantities.
Before turning to the main subject of my talk, the computation of form
factors for this theories, I shall briefly review some of their properties
which will be of relevance. I shall start with the classical theory of which
many features survive in the quantum version. It is most conveniently
described by the Lagrangian density
\beq
{\cal L} = tr \left( \frac{1}{2} \partial_{\mu} \Phi \partial^{\mu} \Phi -
e^{\beta \Phi} E e^{- \beta \Phi} E^{\dag} \right) \;\; , \label{eq: lagra}
\eeq
where $\Phi$ is an $x,t$ dependent field contained in the Cartan subalgebra
of the simple Lie algebra {\bf g} underlying the theory, E is a cyclic
element of {\bf g} and the coupling constant $\beta$ is taken to be real.
Developing the above Lagrangian to second order in $\Phi$ it has been
proven \cite{FLO} that the masses are proportional to the right
Perron-Frobenius
vector $y_i$ of the Cartan matrix
\beq
m_i = \beta y_i \sin \frac{\pi}{h}  \;\;  ,
\eeq
with h denoting the Coxeter number of {\bf g}. Going to the next order one
finds that the three point couplings are proportional to the area bounded
by the sides of length equal to the masses of the fusing particles,
$\Delta_{ijk}$
\beq
C_{ijk} = \frac{4 i \beta \varepsilon(i,j,q)}{ \sqrt{h}} \Delta_{ijk}
\eeq
where $\varepsilon(i,j,q)$
\footnote{Denoting by $\Psi^2$ the length of the highest root, $\sqrt{2/\Psi^2}
\varepsilon(i,j,q)$ acquires always the value $\pm 1$, unless all three
particles correspond to short roots, in which case it becomes $\pm 1/ \sqrt{2}$
for $B,C,F_4$ and $\pm 2/\sqrt{3}$ for $G_2$.}
denotes a structure constant of the algebra. A selection rule which decides
whether the coupling constant is non-vanishing can be cast elegantly in the
root
space of {\bf g}, i.e. $C_{ijk} \neq 0$ if and only if
\beq
\sigma^{\xi(i)} \gamma_i + \sigma^{\xi(j)} \gamma_j +\sigma^{\xi(k)} \gamma_k
= 0 \;\;   .\label{eq: frule}
\eeq
Here $\sigma$ denotes the Coxeter transformation, the triplet of integers
$(\xi(i), \xi(j), \xi(k) )$ form an equivalence class in which each integer
takes values between $0$ and $h$ and $\gamma_i$ a simple root $\alpha_i$
associa
   ted
to particle $i$ multiplied by the colour value $c(i)=\pm 1$, due to the
bicolouration of the Dynkin diagram. Going to higher order one finds that
all n-point couplings with $n \geq 4$ can be completely determined in terms
of the masses and the three-point couplings \cite{AF}. A generating function
for all couplings is given by
\beq
C_{l_{1} \dots l_{n}} =  (-1)^{n+1} \sum_{t=1}^{n-2} \sum\limits_{
\leftrightarrow \atop x}
\frac{x_1 \dots x_t}{{\cal N}_t} m_{l_{n-1}}^2 \delta_{\bar{l}_{n-1}  l_{n}}
  \;\; ,\label{eq: multipoint}
\eeq
where the quantities $x_i$ for each particular $t$ are expressible in terms of
the three point couplings $x_1, \dots, x_{2(t+1) -n} = \sum_{\kb} \frac{C_{\mu
\
   nu \kb}}{m_k^2} $ or the masses $x_{2t +3 -n}, \dots, x_t = m_{l{\nu}}^2
\delta_{\bar{l}_{\nu}  l_{\mu}} $ .
Here $\sum\limits_{\leftrightarrow \atop x}$ denotes the sum over all possible
permutations of the $x_i$. The factor ${\cal N}_t$ takes care of the
overcountin
   g
when symmetric terms are permuted. Its explicit value is given by
$ {\cal N}_t = (2t +2 -n) ! (n - t -2) !  $  .

The central object for the on-shell quantum theory, the two-body scattering
matrix, can be expressed very compactly for ATFT related to simply laced
algebras
\beq
S_{ij}(\th) = \prod_{q=1}^{h} \left\{ 2q - \frac{ c(i) + c(j)}{2}
\right\}_{\th}^{- \frac{1}{2} \lambda_i \cdot \gamma^q \gamma_j } \;\;\; .
\label{eq: smat}
\eeq
Here $\th$ denotes the rapidity, $\lambda_i$ a fundamental weight of the
algebra and $\{ \}_{\th}$ is a building block consisting out of sinh-functions,
i.e. $\{ x\}_{\th} = [ x ]_{\th}/ [x]_{-\th}, [x]_{\th} = < x+1>_{\th}
<x-1>_{\th} / < x+1-B>_{\th} <x-1+B>_{\th} $ and $< x>_{\th} = \sinh\frac{1}{2}
\left( \th + \frac{i \pi x}{h}\right)$. $B(\beta)$ is a function containing the
coupling constant dependence which takes values between 0 and 2. The S-matrix
possess the remarkable property to be left invariant when mapped as $B
\rightarrow 2 - B$, i.e. the theory contains a duality between the strong
and weak coupling region. Formula (\ref{eq: smat}) has been found to satisfy
all
the consistency requirements demanded from a scattering matrix \cite{FO} and
is furthermore supported by perturbative checks \cite{BS}. A particular nice
example where some classical features survive in the quantum theory is the
fusin
   g
rule (\ref{eq: frule}) which becomes equivalent to the so-called {\it bootstrap
equation}.
\beq
S_{li}\left( \th + i \eta(i) \right) \;S_{lj}\left( \th + i \eta(j) \right) \;
S_{lk}\left( \th + i \eta(k) \right) \; = \; 1
\eeq
with $ \eta(t) = - \frac{\pi}{h} \left( 2 \xi(t) + \frac{1 - c(t)}{2} \right)$
for $t=i,j,k$.

 Once the on-shell physics is understood, the question towards the
off-shell physics arises naturally and it turns out that the knowledge of
the S-matrix is useful to compute off-shell quantities like correlation
functions of some operator, say $\cal{ O}$
\beq
{\cal G}_n (r_1, \dots r_n) = \langle {\cal O}_1(r_1)\dots{\cal
O}_n(r_n)\rangle
\eeq
where $r$ is denoting the radial distance $r = \sqrt{ x_{0}^2 + x_{1}^2 }$. In
particular the two point function for an hermitian operator ${\cal O}$
in real Eucledian space is given by
\bea
& &\langle{\cal O}(x)\,{\cal O}(0)\rangle\,=
\sum_{n=0}^{\infty}
\int \frac{d\th_1\ldots d\th_n}{n! (2\pi)^n}
 \langle 0|{\cal O}(x)|\th_1,\ldots,\th_n \rangle_{\rm in}{}_{\rm in}
 \langle \th_1,\ldots,\th_n|{\cal O}(0)|0 \rangle \nonumber \\
& &\hspace{3mm} =\,\sum_{n=0}^{\infty}
\int \frac{d\th_1\ldots d\th_n}{n! (2\pi)^n}
\mid F_n^{\cal O}(\th_1\ldots \th_n)\mid^2 \exp \left(-mr\sum_{i=1}^n\cosh\th_i
\right) , \label{eq: twopoint}
\eea
where the so-called {\it form factors} of an operator ${\cal O}$ have been
intro
   duced
\beq
F_n^{\cal O} (\th_1, \dots , \th_n) = \langle 0 |  {\cal O} |\th_1, \dots ,
\th_
   n
\rangle_{\rm in}  \;\; .
\eeq
Thus once one has obtained all n-particle form factors, one is left with
the problem to compute the sum and the integrals in (\ref{eq: twopoint})
in order to obtain two-point correlation functions. In comparison with a
perturbative expansion in the coupling constant, this approach has the
advantage that it contains the coupling constant dependence to all orders.
Furthermore, the integrals and the sum are fast convergent. In the rest of
my talk I shall be concerned with the explicit computation of form factors
$F_n^{\cal O} (\th_1, \dots, \th_n)$. It is based on a sequence of papers
jointly written with G. Mussardo and P. Simonetti \cite{SINH,FMS},
to whom I'm grateful for collaboration.

\section{General Properties of Form Factors}

I shall now state in a fairly axiomatic fashion some general properties which
are expected to hold for form factors. Due to lack of time and space I shall
refer for a proper justifications  of them in terms of general principles of
quantum field theory to various places in the literature [7-14]. In order to
keep the notation as simple as possible I shall concentrate on theories
involving solely one type of particle, since it will be for this theories
where I present a solution of the axiomatic system in the next section.

\noindent
{\bf i) \underline{Commutation of States}}

As a consequence of the commutation of two operators in the Zamolodchikov
algebra, one obtains
\beq
F_n^{\cal O} ( \th_1, \dots , \th_i \th_{i+1}, \dots \th_n ) =F_n^{\cal O}
( \th_1, \dots , \th_{i+1} \th_i, \dots \th_n ) S_{i (i+1)}(\th_{i (i+1)}) \;\;
   .
\label{eq: watson1}
\eeq

\noindent
{\bf ii) \underline{Analytic Continuation}}

The discontinuity at the cuts at $\th_{ij} = 2 \pi i$ is fixed by
\beq
F_n^{\cal O}(\th_1 + 2 \pi i, \th_2, \dots,\th_n) = F_n^{\cal O}
(\th_2, \dots, \th_n,\th_1) = \prod_{l=2}^n S_{1l}( \th_l - \th_1)
F_n^{\cal O}( \th_1, \dots , \th_n) \;\;  . \label{eq: watson2}
\eeq

\noindent
{\bf  iii) \underline{Relativistic Invariance}}

Since we are describing relativistic invariant theories, we expect for an
operat
   or
{\cal O} with spin $s$
\beq
F_n^{\cal O}( \th_1+ \Delta, \dots , \th_n + \Delta) = e^{s \Delta} \;
F_n^{\cal O}( \th_1, \dots , \th_n) \;\; .
\eeq

\noindent
{\bf iv) \underline{Kinematic Residue Equation}}

Describing theories with equal masses we have a kinematical pole at $i \pi$
(for different masses we require (\ref{eq: multipoint})), which leads to
a recursive equation connecting the $(n+2)$ and $n$-particle form factor
\beq
-i \lim_{ \bar{\th} \rightarrow \th} ( \bar{\th} - \th ) F_{n+2}^{\cal O}
( \bar{\th} + i \pi, \th , \th_1, \dots, \th_n) \; = \; \left(
1-\prod_{i=1}^{n}
S( \th - \th_i) \right)  F_{n}^{\cal O}(  \th_1, \dots, \th_n)
\label{eq: kinres}
\eeq

\noindent
{\bf v) \underline{Bound State Residue Equation}}

A further pole arises as a consequence of a possible bound state, due to the
process $ i + j \rightarrow  k $, in which case we obtain a recursive
equation connecting the $(n+1)$ and the $n$-particle form factors
\beq
-i \lim_{ \bar{\th} \rightarrow \th} ( \bar{\th} - \th ) F_{n+1}^{\cal O}
( \bar{\th} + i \eta(i), \th + i \eta(j), \th_1, \dots, \th_n) \; = \;
\Gamma_{ij}^{k}  F_{n}^{\cal O}(  \th + i \eta(k), \th_1, \dots, \th_n) \; ,
\label{eq: boundres}
\eeq
where $\Gamma_{ij}^{k}$ denotes the three particle vertex on mass-shell.
\vfill \break
\noindent
{\bf  vi) \underline{Cluster Formula}}

An interesting equation results if one shifts part of the operators forming the
physical state to infinity
\beq
\lim_{\Delta \rightarrow \infty} F_{k+l}^{\cal O} (\th_1, \dots , \th_k,
\th_{k+1} + \Delta , \dots , \th_{k+l} + \Delta ) F_0^{\cal O}  =
F_k^{\cal O} ( \th_1, \dots , \th_k) F_l^{\cal O}( \th_{k+1}, \dots
\th_{k+l}  ) \; .  \label{eq: cluster}
\eeq

\noindent
{\bf vii) \underline{Form Factors of the Energy Momentum Tensor}}

Once one posses a conservation law involving different kinds of operators one
can always utilize it to relate their form factors. The energy momentum tensor
is particular good example for this since it is present in every theory.
Denoting the trace of the energy momentum tensor by $\Theta$ and the parts
which correspond to the holomorphic and anti-holomorphic in the limit towards
the Eucledian version of a conformal field theory by $T$ and $\bar{T}$,
respectively, we have
\beq
\partial_{\bar{z}} T(z,\bar{z}) + \partial_z \Theta(z,\bar{z}) =  0
\qquad \hbox{and} \qquad \partial_z \bar{T}(z,\bar{z}) + \partial_{\bar{z}}
\Theta(z,\bar{z}) =  0 \;\; ,
\eeq
from which we derive
\footnote{The function $\sigma_k^{(n)}(x_1, \dots, x_n)$ denotes {\it
elementary
symmetric polynomials} which can be generated by
$\prod\limits_{i=1}^n(x+x_i)\,=
   \,\sum_{k=0}^n x^{n-k}\,\sigma_k^{(n)}(x_1,x_2,
\ldots,x_n)$ .}
\bea
\sigma^{(2n)}_1 \sigma^{(2n)}_{2n} F_{2n}^T (\be_1, \dots, \be_{2n} ) &=&
\sigma^{(2n)}_{2n-1} F_{2n}^{\Theta} (\be_1, \dots, \be_{2n} ) \\
\sigma^{(2n)}_{2n-1} F_{2n}^{\bar{T}} (\be_1, \dots, \be_{2n} ) &=&
\sigma^{(2n)}_1 \sigma^{(2n)}_{2n} F_{2n}^{\Theta} (\be_1, \dots, \be_{2n} ) .
\eea

\noindent
{\bf viii) \underline{Form Factors of Descendent Operators}}

Employing the Lorentz group to decompose the space of form factors into
\beq
{\cal P} = \bigoplus_{s} \;\; {\cal P}_s  \quad ,
\eeq
where  $F_n^{\cal O} \in {\cal P}_s$ when ${\cal O}$ has spin $s$, the previous
consistency equations provide a criterion to decide whether a particular
subspace ${\cal P}_s $ is empty or not. Assuming that we can express
an operator ${\cal O'} \in {\cal P}_s$ in the form ${\cal O'} =[ {\cal O}
Q_s]$,
where ${\cal O}$ is spinless, and employing the eigenvalue equation for the
conserved charges
\beq
Q_s \; |\be_1, \dots, \be_n \rangle = q_n^s( \chi_1^s x_1, \dots ,\chi_n^s x_n)
\; |\be_1, \dots, \be_n \rangle
\eeq
we derive the following relations between the two form factors.
\beq
F_n^{\cal O'}(\be_1, \dots, \be_n)\; = \;q_n^s(\chi_1^s x_1,\dots,\chi_n^s x_n)
F_n^{\cal O}(\be_1, \dots, \be_n) \; .
\eeq
The kinematic- and bound state residue equation for the form factor
$F_n^{\cal O'}$ then lead to the two constraints on the eigenvalues for the
conserved charge
\bea
q_{n+2}^s(-\chi^s x,\chi^s x, \chi_1^s x_1,\dots,\chi_n^s x_ n)
&=& q_{n}^s(\chi_1^s x_1,\dots,\chi_n^s x_n) \label{eq: conkin} \\
q_{n+1}^s(\chi^s_i e^{i\bar{u}^{j}_{ik}},\chi^s_j e^{i\bar{u}^{i}_{jk}},
\chi_1^s x_1,\dots,\chi_{n-1}^s x_{ n -1} ) &=& q_{n}^s(\chi^s_k x,
\chi_1^s x_1,\dots,\chi_{n-1}^s x_{ n -1} ) \;\; . \label{eq: conbound}
\eea
Due to the relativistic invariance the eigenvalues will be of the form
\beq
q_{n}^s(\chi_1^s x_1,\dots,\chi_n^s x_n) = \sum_{i=1}^n \chi_i^s e^{s \be_{i}}
= s_k^{(n)} (\chi_1^s x_1,\dots,\chi_n^s x_n)  \;\; .
\eeq
 The polynomial $s_k^{(n)}$ can always be expressed entirely in terms of the
pol
   ynomial $I^{(n)}_{2s-1} = (-1)^{s+1} \det {\cal I}$ ( ${\cal I}$  being
an $(s\times s)$-matrix, whose entries are ${\cal I}_{1j} = \sigma_{2j-1} $
and  ${\cal I}_{ij} = \sigma_{2j-2i+2})$ via equation
\beq
s_k^{(n)} =  \left( I_1^{(n)} \right)^k + k \sum_{\{ \lambda \} } \prod_i
I_{\lambda_{i}}^{(n)} \;\;  , \label{eq: diffb}
\eeq
where $ \{ \lambda \}$ denotes a partition of the integers $\lambda$ into
odd integers, i.e. $ \sum_i \lambda_i = \lambda$ for all $\lambda_i$ odd.
This guarantees that equation (\ref{eq: conkin}) will
always be satisfied due to the invariance property $I^{(n+2)}_{2s-1}  =
I^{(n)
   }_{2s-1}$. On the other hand equation (\ref{eq: conbound}) leads to
the non-trivial condition
\beq
\left( \chi_i^s \right)^s e^{i\bar{u}^{j}_{ik}} + \left( \chi_j^s \right)^s
e^{-i\bar{u}^{i}_{jk}}  = \left( \chi_k^s \right)^s,
\eeq
which is precisely the consistency equation derived by Zamolodchikov
\cite{Zam} in a different context. It selects out a particular set of
possible spin values for descendent operators.

\noindent
{\bf  ix) \underline{Minimal Form Factors}}

In \cite{BKW} it was proven that a general form factor can always be decomposed
into the form
\beq
F_n^{\cal O}( \th_1, \dots , \th_n) \;  = \; K_n^{\cal O}(\th_1, \dots, \th_n)
\prod_{i < j} F_{\rm min} (\th_{ij})
\eeq
where $ K_n^{\cal O} ( \th_1, \dots, \th_n) $ is a totally symmetric function
in $\th_i$, $2 \pi i$ periodic, containing the entire pole structure and
determines the asymptotic behaviour for large values of the rapidities.
On the other hand the minimal form factor contains no poles and zeros in the
phy
   sical sheet,
converges to a constant for $\th_i \rightarrow \infty$ and satisfies
Watson's equations (\ref{eq: watson1}) and (\ref{eq: watson2}) for $n=2$
\beq
F_{\rm min}(\th) \; = \; F_{\rm min} ( - \th) S( \th) \qquad \hbox{and}
\qquad  F_{\rm min} ( i \pi - \th) \; = \; F_{\rm min} (i \pi + \th) \;\;  .
\label{eq: minform}
\eeq

\section{ Form Factors for one scalar field ATFTs}

In this section I shall present a general solution for ATFTs involving one
scalar field only, i.e. the Sinh-Gordon theory and the Bullough-Dodd model.
For the purpose of this talk it will be sufficient to illustrate the general
tec
   hniques involved.

\subsection{The Sinh-Gordon theory}

The Lagrangian density for the Sinh-Gordon theory reads
\beq
{\cal L}\,=\, \frac{1}{2}(\partial_{\mu}\phi)^2-\frac{m^2}{g^2}
\,\cosh\,g\phi(x) \,\,\, .
\eeq
The theory posses a $Z \!\!\!\! Z_2$-symmetry, that is, it remains invariant
under the map $\phi \rightarrow -\phi $. Its conservation laws are graded by
the spin values $1,3,5,7, \dots $ and its S-matrix is given by $S(\th) =
\{ 1 \}_{\th}$, which does not posses a pole in the physical sheet such that a
fusing process is not possible. With the scattering matrix an input we can
solve (\ref{eq: minform}) by the usual technique of using Fourier transform
after taking the logarithm of this equation and obtain
\beq
F_{\rm min}(\th,B)\,=\,
\prod_{k=0}^{\infty}
\left|
\frac{\Gamma\left(k+\frac{3}{2}+\frac{i\hat\th}{2\pi}\right)
\Gamma\left(k+\frac{1}{2}+\frac{B}{4}+\frac{i\hat\th}{2\pi}\right)
\Gamma\left(k+1-\frac{B}{4}+\frac{i\hat\th}{2\pi}\right)}
{\Gamma\left(k+\frac{1}{2}+\frac{i\hat\th}{2\pi}\right)
\Gamma\left(k+\frac{3}{2}-\frac{B}{4}+\frac{i\hat\th}{2\pi}\right)
\Gamma\left(k+1+\frac{B}{4}+\frac{i\hat\th}{2\pi}\right)}
\right|^2 \;\; ,
\eeq
$( \hat{ \th} = i \pi - \th)$ which satisfies the functional equation
\beq
F_{\rm min} (\th + i \pi, B) F_{\rm min} (\th, B) \; =\; [ 1 ]_{\th} \;\; .
\eeq
This identity is required in the process of solving the recursive
equation resulting from (\ref{eq: kinres}). It turns out to be useful to
parameterize the the non-minimal part of the form factor further and extract
the polestructure from it
\beq
K_n(\th_1, \dots, \th_n) = \frac{Q_n(\th_1, \dots , \th_n)}{ \prod\limits_{i <
j
   }
x_i   + x_j  }   \;\; ,
\eeq
where $x_i = e^{\th_i} $ is introduced. Furthermore, we obtain that the
function

$Q_n (x_1, \dots , x_n)$ can always be written as
\beq
Q_n(x_1, \dots, x_n) = \left\{
\begin{array}{ll}
\sigma_n^{(n)} P_n(x_1,\dots,x_n) & \qquad \hbox{for} \;  \phi  \\
\sigma_1^{(n)}  \sigma_{n-1}^{(n)} P_n(x_1,\dots,x_n) & \qquad \hbox{for} \;
\Theta
\end{array}    \right.  \label{eq: paraq}
\eeq
then the functions $P_n(x_1,\dots,x_n)$ obey the recursive equation
\beq
(-1)^{n+1} P_{n+2}( -x,x, x_1, \dots , x_n) \; = \; x \;
D_n(x,x_1, \dots , x_n)  P_n(x_1, \dots , x_n)
\eeq
with
\beq
D_n={{1} \over {2 \sin(\pi B/2)}} \sum_{l,k=0}^n (-1)^l \sin\left(
(k-l) {{\pi B} \over{2}} \right) x^{2n -l-k} \sigma_l^{(n)} \sigma_k^{(n)}
\eeq
as a result of (\ref{eq: kinres}). Due to the $Z \!\!\!\! Z_2$-symmetry we have
$P_{2n}^{\phi} = P_{2n +1 }^{\Theta}  = 0$. Fixing the initial condition
of the recursive equation to be $F_1^{\phi}(\th_1) = \frac{1}{\sqrt{2}}$ and
$F_2^{\Theta} = 2 \pi m^2$ we can solve this equation iteratively
\bea
P_3(x_1,\ldots,x_3)         &=& 1 \nonumber  \\
P_4(x_1,\ldots,x_4) &=& \sigma_2\nonumber \\
P_5(x_1,\ldots,x_5) &=&  \sigma_2 \sigma_3 - 4 \cos^2\left( \frac{\pi B}{2}
\right) \sigma_5 \\
P_6(x_1,\ldots,x_6) &=& \sigma_2\sigma_3 \sigma_4  + \left( 4 \cos^2\left(
\frac
   {\pi B}{2} \right) -1 \right) \sigma_3  \sigma_6)  - 4 \cos^2\left(
\frac{\pi
    B}{2}\right) (\sigma_4 \sigma_5 + \sigma_1 \sigma_2 \sigma_6) \;\; .
\nonumber
\eea
Obviously we can carry on and compute iteratively $P_n$ for higher values
of n. However if we are solely concerned with the computation of correlation
function, the first few terms are usually sufficient to achieve high
precision. In \cite{SINH} we have verified this with the particular
application to the correlation function of the energy-momentum tensor, which
is expressed via the c-theorem. None-the-less for particular values of the
coupling constant we can prove that the $P_n$ take on relatively simple forms.
For instance  for the self-dual point, i.e. $B(\sqrt{8 \pi}) = 1$ we obtain
\beq
P_n(x_1, \dots , x_n) \; = \;  \det {\cal A}^{(n)}(x_1, \dots , x_n)
\eeq
where ${\cal A}^{(n)}(x_1, \dots , x_n)$ is a $(n-3) \times (n-3)$-matrix whose
   entries are given by
\beq
{\cal A}_{ij}^{(n)}\,=\, \sigma^{(n)}_{2j-i+1}\,\cos^2\left((i-j)\frac{\pi}{2}
\right) \,\,\, .
\eeq
Whereas for the inverse Yang-Lee point, i.e. $B \left( 2 \sqrt{ \pi} \right) =
\frac{2}{3}  $, we obtain
\beq
P_n(x_1, \dots , x_n) \; = \;  \det {\cal B}^{(n)}(x_1, \dots , x_n)
\eeq
where the entries of the matrix are now given by
\beq
{\cal B}_{ij}^{(n)} \; = \; \sigma^{(n)}_{3j -2i +1}  \;\;\; .
\eeq
It is interesting to note that the recursive equation only posses two
solutions,
suggesting that $\phi$ and $\Theta$ are the fundamental operator content of the
theory. As a non-trivial check of our solution it is instructive to verify
that it satisfies the cluster equation (\ref{eq: cluster}). Hence we have found
a solution for all the axioms presented in the previous section.

\subsection{The Bollough-Dodd model}
I shall now present a further example, which involves the additional
complicatio
   n
of the involvement of the bound state residue equation, which was absent in the
Sinh-Gordon theory due to the lack of fusing process. The Bullough-Dodd
Lagrangi
   an density reads
\beq
{\cal L}\,=\,\frac{1}{2}(\partial_{\mu}\varphi)^2-
\frac{m^2_0}{6g^2}\left(2e^{g\varphi}+e^{-2g\varphi}\right)\,\, .
\label{BuD}
\eeq
The model posses conserved quantities graded by the spins $1,5,7,11,13,\dots$
an
   d
its S-matrix is given by $S(\th,B) = \{ 1 \}_{\th} \{ 2 \}_{\th}$ \cite{ZZ},
which possess a pole a $i \pi /3$ representing the fusing process
$ A + A \rightarrow  A$. In the same fashion as in the previous subsection
we solve (\ref{eq: minform}) and find
\begin{eqnarray}
&& F_{\rm min}(\th,B)\,=\,
\prod_{k=0}^{\infty}
\left|
\frac{\Gamma\left(k+\frac{3}{2}+\frac{i\hat\th}{2\pi}\right)
\Gamma\left(k+\frac{7}{6}+\frac{i\hat\th}{2\pi}\right)
\Gamma\left(k+\frac{4}{3}+\frac{i\hat\th}{2\pi}\right)}
{\Gamma\left(k+\frac{1}{2}+\frac{i\hat\th}{2\pi}\right)
\Gamma\left(k+\frac{5}{6}+\frac{i\hat\th}{2\pi}\right)
\Gamma\left(k+\frac{2}{3}+\frac{i\hat\th}{2\pi}\right)}
\right.
\\
&&\,\,\,\times\,\,\left.
\frac{\Gamma\left(k+\frac{5}{6}-\frac{B}{6}+\frac{i\hat\th}{2\pi}\right)
\Gamma\left(k+\frac{1}{2}+\frac{B}{6}+\frac{i\hat\th}{2\pi}\right)
\Gamma\left(k+1-\frac{B}{6}+\frac{i\hat\th}{2\pi}\right)
\Gamma\left(k+\frac{2}{3}+\frac{B}{6}+\frac{i\hat\th}{2\pi}\right)}
{\Gamma\left(k+\frac{7}{6}+\frac{B}{6}+\frac{i\hat\th}{2\pi}\right)
\Gamma\left(k+\frac{3}{2}-\frac{B}{6}+\frac{i\hat\th}{2\pi}\right)
\Gamma\left(k+1+\frac{B}{6}+\frac{i\hat\th}{2\pi}\right)
\Gamma\left(k+\frac{4}{3}-\frac{B}{6}+\frac{i\hat\th}{2\pi}\right)}
\nonumber \right|^2
\end{eqnarray}
which now satisfies the functional equation
\bea
F_{\rm min} (\th + i \pi, B) F_{\rm min}(\th,B) & = &  [1]_{\th} [2]_{\th} \\
F_{\rm min} (\th + i \pi/3, B) F_{\rm min}(\th - i \pi/3,B) & = &  [0]_{\th}
F_{\rm min} (\th , B) \;\; .
\eea
Extracting again the polestructure from the minimal part of $F_n$, we obtain
\beq
F_n(\beta_1,\ldots,\beta_n)\,=\, Q_n(x_1,\ldots,x_n)\,
\prod_{i<j} \frac{F_{\rm min}(\beta_{ij})}{(x_i+x_j)(\omega x_i+x_j)
(\omega^{-1}x_i+x_j)} \,\, .
\eeq
( $\omega = e^{i \pi /3}$) which we further factorize as in (\ref{eq: paraq}).
Finally we derive the recursive equation resulting from the kinematic residue
equation
\beq
(-1)^{n} Q_{n+2}(-x,x,x_1,x_2,\ldots,x_n)\,=
\,\frac{1}{F_{\rm min}(i\pi)}\,x^3 \,U(x,x_1,x_2,\ldots,x_n)
Q_n(x_1,x_2,\ldots,x_n) \label{eq: bdkin}
\eeq
where
\bea
U(x,x_1,\ldots,x_n) &=& 2\sum_{k_1,\ldots,k_6=0}^n
(-1)^{k_2+k_3+k_5}\, x^{6n-(k_1+\cdots\, +k_6)}
 \,\sigma_{k_1}^{(n)}\sigma_{k_2}^{(n)}\ldots\sigma_{k_6}^{(n)} \, \\
&&\,\,\, \times\,
\sin\left[\frac{\pi}{3}\left[2(k_2+k_4-k_1-k_3)+B(k_3+k_6-k_4-k_5)\right]
\right] \nonumber
\eea
and the bound state residue equation
\beq
Q_{n+2}(\omega x,\omega^{-1}x,x_1,\ldots,x_n)\,=\,-\,
\frac{\sqrt{3}}{F_{\rm min}(i \frac{2\pi}{3})} \,
\Gamma(g) \,x^3 D(x,x_1,\ldots,x_n) \,Q_{n+1}(x,x_1,\ldots,x_n)
\label{eq: bdbound}
\eeq
where
\beq
D(x,x_1,\ldots,x_n)\, = \,\sum_{k_1,k_2,k_3=0}^n
x^{3n-(k_1+k_2+k_3)} \, \omega^{(2+B)(k_2-k_3)} \,
 \,\sigma_{k_1}^{(n)}\sigma_{k_2}^{(n)}\sigma_{k_3}^{(n)} \,\,\,.
\eeq
The three-particle vertex on mass-shell acquires the form
\beq
\Gamma^2(B)\,=\,2{\sqrt 3} \frac{\tan\left(\frac{\pi B}{6}\right)}
{\tan\left(\frac{\pi B}{6}-\frac{2\pi}{3}\right)}
\frac{\tan\left(\frac{\pi}{3}-\frac{\pi B}{6}\right)}
{\tan\left(\frac{\pi B}{6}+\frac{\pi}{3}\right)}\,\, . \label{eq: threepv}
\eeq
Apart from vanishing in the free theory for $B=0,2$, this function becomes
zero as well at the self-dual point for $B=1$. This is a peculiar feature not
known in ATFTs related to simply laced algebras. At this particular point the
S-matrix coincides with the one of the Sinh-Gordon theory at the inverse
Yang-Lee point. In \cite{FMS} we showed that this feature extends off-shell,
that is
\beq
F_n^{BD}\left( B=1,x_1, \dots, x_{n} \right) \, = \,
F_n^{SG}\left( B=\frac{2}{3},x_1, \dots, x_n \right) \,\,\, ,
\eeq
for both the elementary field and the energy-momentum tensor. Thus the $Z
\!\!\!
\!Z_2$-symmetry is realised dynamically in the Bullough-Dodd model.
Furthermore,
    we showed that equations (\ref{eq: bdkin}) and (\ref{eq: bdbound}) indeed
po
   sses
consistent solutions in accordance with all consistency requirements.

\section{Conclusions}
We conclude that the form factor approach indeed provides a successful
method to compute correlation functions for massive integrable models.
For particular values of the coupling constant it has been shown that one can
obtain closed formulas for $F_n$. It is desirable to extend this to arbitrary
values of $\beta$ and obtain closed expressions for those values too. Despite
the fact that high numerical precision can be achieved relatively easy in
(\ref{eq: twopoint}), by computing only the first few terms, it is desirable to
obtain a closed analytic solution of it. Alternatively it would be illuminating
to find out whether it is possible to obtain correlation functions via
differential equations satisfied by the form factors.
\par \noindent
{\em Acknowledgement}: I would like to thank G. Mussardo and P. Simonetti for
a fruitful collaboration on the main part of this presentation and R. K\"oberle
for useful discussions and some of the speaking time originally allocated to
him
   .
Furthermore I gratefully acknowledge the support of the Funda\c{c}\~ao de
Amparo \`a Pesquisa do Estado de S\~ao Paulo (FAPESP).

\end{document}